*Chapter 1*

# The Internet of things: a survey and outlook

*Giovanni Perrone[1], Massimo Vecchio[2], Javier Del Ser[3], Fabio Antonelli[2], and Vivart Kapoor[4]*

The recent history has witnessed disruptive advances in disciplines related to information and communication technologies that have laid a rich technological ecosystem for the growth and maturity of latent paradigms in this domain. Among them, sensor networks have evolved from the originally conceived set-up where hundreds of nodes with sensing and actuating functionalities were deployed to capture information from their environment and act accordingly (coining the so-called wireless sensor network concept) to the provision of such functionalities embedded in quotidian objects that communicate and work together to collaboratively accomplish complex tasks based on the information they acquire by sensing the environment. This is nowadays a reality, embracing the original idea of an Internet of things (IoT) forged in the late twentieth century, yet featuring unprecedented scales, capabilities and applications ignited by new radio interfaces, communication protocols and intelligent data-based models. This chapter examines the latest findings reported in the literature around these topics, with a clear focus on IoT communications, protocols and platforms, towards ultimately identifying opportunities and trends that will be at the forefront of IoT-related research in the near future.

## 1.1 Introduction

The term 'Internet of things' (IoT), coined in 1999 by Ashton [1], has attracted and attracts a multitude of research and industrial interests. Whatever domain we take into consideration, from housing to precision agriculture, from retail to transportation, from infrastructure monitoring to personal healthcare, from urban mobility to autonomous vehicles, just to mention a few, is going to be supported by each-day-smaller and smarter devices (i.e., things) able to collect data and to push them to the Internet. Gartner Inc. expects that by 2020, 26 billion objects will be connected to the Internet and, by 2022, a typical family home will contain more than 500 connected smart objects [2].

[1]SMARTEST Research Centre at eCampus University, Novedrate (Como), Italy
[2]OpenIoT Research Unit at FBK CREATE-NET, Trento, Italy
[3]ICT Division, TECNALIA, Derio, Bizkaia, Spain; Department of Communications Engineering, University of the Basque Country (UPV/EHU), Bilbao, Spain; Data Science group, Basque Center for Applied Mathematics (BCAM), Bilbao, Spain
[4]Endress+Hauser Process Solution GmbH, Dept. IoT DE, Freiburg, Germany

**K. Ashton says:**

*If we had computers that knew everything there was to know about things using data they gathered without any help from us we would be able to track and count everything, and greatly reduce waste, loss and cost. We would know when things needed replacing, repairing or recalling, and whether they were fresh or past their best. We need to empower computers with their own means of gathering information, so they can see, hear and smell the world for themselves, in all its random glory. RFID and sensor technology enable computers to observe, identify and understand the world without the limitations of human-entered data.* [1]

However, around 20 years after its coinage [3], it would be reductive to translate the IoT vision into the routine process of physically connecting everyday objects to the Internet. Indeed, the original promise of the IoT was about connecting such everyday objects to the Internet with the goal of making them autonomous, hence smart. Loosely speaking, this smartening process relies on the implied assumption of the need of 'closing the loop' that consists of sensing, communicating, reasoning, taking decisions and actuating back into the physical side [4]. As already noticed by the authors of [5], this can be seen as the specification of the more general Monitor-Analyse-Plan-Execute plus Knowledge (MAPE-K) reference model for Autonomic Computing, as initially proposed by Horn as far back as 2001, and later formalized in [6] (the interested reader is also invited to refer to [7]).

**P. Horn says:**

*It's time to design and build computing systems capable of running themselves, adjusting to varying circumstances, and preparing their resources to handle most efficiently the workloads we put upon them. These autonomic systems must anticipate needs and allow users to concentrate on what they want to accomplish rather than figuring how to rig the computing systems to get them there [ . . . ] it is the self-governing operation of the entire system, and not just parts of it, that delivers the ultimate benefit.* [8]

Overall, the number of devices connected to the Internet is exploding, as well as the volume of data produced and pushed to the Internet by these devices. To support this uncontrolled proliferation of physical devices and associated, hence tremendous, digital data production, IoT embraces a number of architecture, protocols, standards, services and applications for ubiquitous data acquisition (at the physical level), as well as large-scale data storage and analysis tools (at the digital level). On the one hand, this guarantees a genuine evolution of the IoT towards a truly open, highly pervasive and multivendor ecosystem. On the other hand, too often we observe how this lack of well-established, horizontal, truly scalable and consistent IoT architecture translates into the sharp tendency of each new application (even worse in some cases: each new device) that enters the IoT arena to bring its own (sometimes proprietary!) technical implementations and manuals. It is clear that this tendency is one of the causes of the today IoT market fragmentation. The obvious – though incumbent – statement is that the IoT market is fragmented because of its very nature. Indeed, very specific end-users' requirements cannot be always homogenized. This is one of the reasons why, sometimes, even apparently simple IoT use cases are built on top of complex system architecture that cannot be promoted as the reference one [9]. However, the whole truth is that major counterforces are slowing down the exploitation of the full potential of the IoT. The most critical challenge that the IoT has tackled since the beginning – and not completely defeated, yet! – is this sort of mistrust shown by well-established large players: the IoT implementation may involve radical structural changes and drastic shift in value creation. It follows that for such players, it is difficult to quickly adapt to new business models and engage in new types of alliances. This inertia explains also why Small and Medium Enterprises (SMEs – esp. entrepreneurs and start-ups) are today considered to have the full potential to seize new opportunities brought up by the IoT. Nevertheless, to have such many actors acting unilaterally promotes the proliferation of proprietary system architecture and technology, as well as business models. To complicate things in this Babylon, we still certify the lack of common standards and interoperable solutions throughout the products and services life cycles, while interoperability is still considered essential to ensure seamless flow of data across sectors and value chains. In the end, this reality does not encourage cross-cutting approaches, risks reinforcing silos, and prevents innovation across areas. However, the opposite situation would be also undesirable: in the end, monopolizing the IoT would be then an insurmountable obstacle to the development of these markets, and to the development of truly open digital platforms.

Thus, it is not a coincidence the current proliferation of such many platforms, standards and technologies claiming to be able to work across multiple vendors, industries and sectors, as well as such a multitude of SMEs focusing efforts on very specific vertical segment use cases and applications [10]. The objective of this chapter is to take stock of the present of the IoT, starting from the enabling radio technologies (see Section 1.2), through the most common communication protocols (see Section 1.3), until the state-of-the-art backend and frontend services and features provided by the modern IoT platforms available on the market and within the open-source communities. Finally, Section 1.5 concludes this chapter by outlining research challenges related to IoT communications, protocols and services that are throwing up manifold questions in need of further investigation. We will argue that these challenges, which lie at the core of foreseen evolutions of IoT architecture and systems, serve as a stimulating incentive and a propelling driver for research efforts conducted in this vibrant field in the future.

## 1.2    Communication and transport technologies

An essential requirement for implementing a large and cooperating network of devices is the ability to have a flexible, inexpensive and adaptive connectivity layer. The restrictions and limitations of the physical cabling, both in terms of implementation costs and lack of flexibility in adapting to the ever-changing dynamics of the environment being monitored, make fixed cabling a solution that, in practical terms, is not viable for several IoT scenarios. Not to mention applications where the devices to be connected are moving, or are simply too far away from each other to be physically connected by wires. For these reasons, during the years, several wireless connectivity standards have emerged, ranging from very short-range, low-power connections to long-range connectivity solutions based on cellular technologies. This section will provide an overview of some of the most common and widely adopted radio technologies used for IoT implementations, emphasizing advantages and disadvantages of each protocol for typical IoT scenarios.

### *1.2.1    Short range*

Short-range solutions are aimed at creating and managing networks of devices that are physically located in a confined area. Such networks are typically called Wireless Personal Area Network, also to emphasize the fact that they operate within a short radius and are separated from other networks. We will see that it is possible to partially overcome the limitations related to the distance between the nodes using meshed and cluster-tree networks (i.e., a sort of cascade of networks connected together) but in any event two elements belonging to the same network must be within a few hundred metres of distance (maximum) in order for connectivity to be established. While short range may seem a disadvantage at first sight, for several applications (e.g., smart home), a maximum connectivity range of 200 metres can be more than enough. To counterbalance the limitations on range, very low power consumptions (directly corresponding to long battery lifetime) and relatively easy protocol stacks allow implementation of these standards on small and inexpensive devices.

#### 1.2.1.1    802.15.4

The 802.15.4 standard has been created and is maintained by the IEEE WPAN Task Group 4, which was chartered to investigate a low data rate solution with multi-month to multiyear battery life and very low complexity. It is operating in an unlicensed, international frequency band. Potential applications are sensors, interactive toys, smart badges, remote controls and home automation. The previous definition embeds a large part of the most important characteristics of the 802.15.4, which can be summarized as follows:

- low data rate (maximum 250 kbps);
- low power consumption;
- uses unlicensed bands;
- allows implementation in very simple devices (e.g., microcontrollers);

- suitable for timing critical applications (access to the network in maximum 30 ms) and
- supports star, meshed and cluster-tree networks.

The standard defines only the first two levels of the OSI protocol stack (i.e., the Physical and the MAC layers), leaving to others to define how to manage network addressing, transport, flow control (if any) and so on. For these reasons, on top of 802.15.4, a number of different standards have been devised in order to implement specific functionalities and features.

A network based on 802.15.4 can be composed of two types of devices: 'end devices' and 'full function devices' (FFD) that allow the creation of meshed and cluster-tree types of networks. The possibility of connecting multiple FFD together allows for a great degree of freedom in shaping dynamic networks that can grow (and shrink) based on the specific needs of the application.

Physical layer uses channels in the unlicensed (but regulated) frequencies for Industrial, Scientific and Medic (ISM) bands (2.4 GHz worldwide, 868 MHz in Europe and 915 MHz in the Americas), allowing for a maximum data rate of 250 kbps if the 2.4 GHz band is used. BPSK or O-QPSK modulations are used. The media access control (MAC) layer takes care of addressing, collision avoidance and access to transmission channels as well as security. Security is implemented through 128-bit AES symmetric cryptography (keys can be specified by the upper layers of the protocol stack).

### 1.2.1.2   ZigBee

ZigBee offers a full protocol stack that can be easily implemented in small, battery-operated devices to create WPANs – the protocol has been designed and is maintained by the ZigBee Alliance, a nonprofit organization that has more than 150 companies, including big names like GE, Huawei, Amazon, etc. ZigBee is based on 802.15.4 and implements the Physical and MAC layers, as specified in the IEEE standard. As such, ZigBee is able to reach a maximum of 250 kbps data rate; considering however that the typical use cases for a ZigBee network require the exchange of payloads of a few hundred bytes, it is easy to realize that the limited data rate is not actually an issue.

One of the key concepts of ZigBee is the Key-Value Pair (KVP): similarly to what happens in publish-subscribe protocols, information is distributed within the network based on the contents to be communicated (and not on a possible list of recipients). To give an example, this means that, when an end node wants to notify a change of state of one of its sensors, it will notify the network coordinator of a change for that specific KVP associated to that sensor. The coordinator will know how to distribute this information based on binding tables, that is, lists of end nodes that have subscribed to receive updates for that KVP. ZigBee protocol stacks add, on top of the Physical and MAC layers, two more layers as follows:

- Network layer: it takes care of addition/removal of nodes into a network and routing of packets between the nodes. The routing algorithm is ad hoc on-demand distance vector (AODV) for meshed networks, while for the other types of networks only static routing is allowed. It is interesting to note that meshed networks, complemented by AODV routing, improve the reliability and resilience of the ZigBee networks, allowing dynamic routing of packets even when one or more nodes are not available and

● Application Framework layer: it is responsible for maintaining the KVPs and for the discovery functions of networks. The application framework (AF) layer implements also the ZigBee define object (ZDO), that is, a software object that provides access to a number of functions used to manage the ZigBee network by the application. One of the biggest advantages of ZigBee is the wide availability of plug-in modules that can be easily connected to microcontrollers and similar small devices, effortlessly adding wireless capabilities with limited expenses.

### 1.2.1.3   Bluetooth low energy

Bluetooth low energy (BLE), sometimes also called Bluetooth Smart, is a WPAN standard based on Bluetooth (BT), and specifically designed to meet the reduced power requirements of those IoT devices equipped with reduced energy availabilities. BLE is defined in release 4.0 of the Bluetooth specifications and uses the same 2.4 GHz radio frequencies of the previous Bluetooth releases, even if the two technologies are not compatible with each other (i.e., a BLE device is not BT-compatible).

BLE offers higher data rate than 802.15.4-based protocols, giving to the end applications the possibility of accessing to bursts of up to 1 Mbps, with however a reduced coverage radius when compared to other solutions. BLE theoretical coverage is, in fact, typically under 100 m, but in real-life typical coverage drops to a few tenths of metres. BLE implements a mesh type of network, that can be used to implement also other network topologies. One of the key advantages of BLE against other protocols is the fact that it is already supported by almost all modern mobile devices; this means that by using BLE it is relatively easy to add support for mobile-based interactions.

Similarly to what we have seen for the 802.15.4-based protocols, devices included in a BLE network may have different roles; such roles are defined in the generic access profile (GAP) and define a distinction between peripheral devices (typically the constrained things to connect) and central devices (typically mobile devices or, however, devices with more processing power than typical peripheral devices).

On top of the already mentioned network topologies supported by other protocols such as point-to-point, star and mesh, BLE supports also a Beacon type of network, where a device sends periodic advertisement messages to let other devices (typically mobile phones) know that they are in proximity of that beacon. This is the technology used, for instance, by Apple iBeacon service and that can allow implementing presence detection, targeted marketing and other applications where it is important to pinpoint with accuracy the indoor location of the user.

### 1.2.1.4   Z-Wave

Z-Wave is a protocol originally developed by a company called Zensys, in 2001, whose specifications have been later released to the public domain (Zensys donated the specifications to ITU who released them in recommendation G.9959 in 2015). It has been designed from the beginning to provide a low data rate wireless communications protocol for home automation applications, creating therefore what is called by the specifications a 'home area network'. Z-Wave is now maintained by Sigma Designs (who acquired Zensys) and the Z-Wave Alliance, the latter is the entity that administers the Z-Wave Standard while the former is the subject that releases interoperability certificates – in fact, interoperability for commercial products is a mandatory requirement for all products that want to support the Z-Wave standard.

The PHY and MAC layers defined in Z-Wave are equivalent, from a logical standpoint, to IEEE 802.15.4. The radio frequencies used are the unlicensed ISM 868–915 MHz spectrum; actual frequencies vary based on the region of operation and, based on the region, different RF profiles (defining data rate and data accuracy values) may be available.

Z-Wave supports data rate up to 100 kbps for the highest performance R3 profile, while the R2 and R1 profiles define data rate of 40 kbps and 9.6 kbps, respectively. Reported average indoor coverage of approx 10 m – this apparent limitation can be overcome using one of the key functionalities sported by Z-Wave, that is, the capability of automatically creating meshed networks. Using this capability, a node can use other nodes (up to four) to transfer a message to another node that cannot be reached directly (either because of distance larger than the maximum allowed coverage, or because of multipath interface or radio holes phenomena). The maximum number of devices per each network is limited to 232 units, but it is possible to connect nodes belonging to different networks (called domains' in the ITU recommendation) using interdomain bridges, therefore actually extending the coverage of a single domain (Table 1.1).

*Table 1.1   A summary of the short-range radio technologies analysed until now. Caution should be taken when analysing data regarding coverage and data rates since the values reported are the best possible ones, which obviously are significantly influenced by several factors (e.g., interference from other radio sources, especially for systems operating at 2.4 GHz)*

| Protocol | PHY | Band | Data rates | Coverage | Max num. devices | Topology | Notes |
|---|---|---|---|---|---|---|---|
| ZigBee | IEEE | 2.4 GHz ISM | 250 Kbps | 20 m | 65,000 | Star, mesh and cluster-tree | Radio modules widely available |
| | 802.15.4 | 868–915 MHz ISM | | | | | |
| BLE | BSIG proprietary | 2.4 GHz ISM | 1 Mbps | 10 m | unlimited | Mesh | Not compatible with normal BT[1] |
| Z-Wave | ITU 9959 | 868–915 MHz ISM | 100 Kbps | 100 m | 232 | Mesh | Proprietary |

[1]But typically implemented in mobile phones.

## *1.2.2   Long range*

While short-range protocols can be successfully used for a number of applications (e.g., smart homes, wearables, etc.), for many other applications larger coverage radii become mandatory. For this reason, several protocols to implement wide area networks (WAN) using different radio technologies have spread in the past few years, implementing different specifications and technologies, but with the basic key requirements of low power consumptions, relatively low data rates and large coverage. Within this framework, an important role could be played in the near future (i.e., starting from 2018) by the introduction of IoT-specific cellular standards that could potentially take advantage of the existing cellular networks infrastructure and equipment.

### 1.2.2.1   Low-power wide area networks (LPWANs)

*Long range (LoRa) and long-range wide area network (LoRaWAN)*

Long range (LoRa) and long-range wide area network (LoRaWAN) are two protocols commonly used in IoT applications, whose luck is due to a mixture of aspects, one of them certainly being the fact that everyone can create a LoRa-based network.

The name LoRa refers to a proprietary PHY layer developed in 2008 by a French company called Cycleo, later acquired by Semtech, and that is used by a number of other higher-level protocols to implement LPWANs. The complement of LoRa, developed and maintained by the LoRa Alliance, is LoRaWAN, a standard that specifies the MAC level and that implements the functionalities required to set up a network and manage accesses to the physical level.

While it is possible to use the LoRa PHY layer with other protocols, the most common combination is LoRa with LoRaWAN and, for this reason, the two are often – erroneously! – used as synonyms.

LoRa uses unlicensed frequencies in the ISM spectrum (867–869 MHz in Europe and 902–328 MHz in the USA) together with adaptive spread spectrum technology based on chirped-FM modulation. The frequency used for the spread spectrum modulation is adapted based on the retransmission rate in order to automatically adjust radio parameters based on the actual error rates. The technology allows enabling longer coverage ranges and at the expenses of data rates. In fact, data rates for LoRa may range between 250 bps and 50 kbps and coverage may reach several kilometres.

The LoRaWAN MAC layer specifies the network topology, that in this case is a star of star, with each node communicating with multiple gateways that are communicating with a network server; interestingly enough, mesh networks are not supported, and therefore it is not possible for an end node to relay messages to other end nodes. In a LoRaWAN network three different types of elements can be found:

- end nodes: typically sensors and endpoints implemented on constrained devices. End nodes can be of three different classes (A, B and C) with different performances, in terms of data rates and latencies made available by the network and, correspondingly, different battery durations;

- concentrators/gateways: acting as transparent bridges between the end nodes and the network server. Connection with the network server is implemented typically through secure TCP/IP links; application data coming from the end nodes is encrypted as well and since concentrators do not implement higher-level protocols, they only act as a pass-through device and
- network server: it is where the actual receiving point of the application is implemented – before reaching the application point packets from concentrators are analysed and possible duplicates discarded.

One last interesting feature worth to mention is the recent introduction of device location capabilities without the utilization of GPS (that may have a consistent impact on the battery duration), notwithstanding the unquestionable advantages that such a functionality would introduce, there are doubts about its real effectiveness, since location of a device on radio signals (e.g., triangulation of the position using distance and signal strength) has proven several times to be extremely complex to obtain with good (or even usable) precision (e.g., below 500–1,000 metres).

The combination of LoRa and LoRaWAN is used in several IoT applications and one of the advantages is that, even if the technology used for the PHY layer is patented, LoRaWAN is an open-source standard that is freely available to whoever subscribes to the LoRa alliance. Therefore, it is possible for everybody to set up and maintain an independent LoRa-based network, without the need for specific licences from Semtech or from the LoRa alliance.

### Sigfox

SigFox is the name of a proprietary technology owned by SigFox, a French company that in 2009 started to work in the LPWA area with a peculiar business model. While, in fact, other solutions like LoRa enable users to deploy and build private LPWA network, SigFox business model is closer to what a typical communication service provider (CSP) or mobile network operator (MNO) would do, leaving SigFox as the only owner of the networks being deployed all over the world. So, while it is possible to buy SigFox-compatible modules at a very low price (SigFox-compatible modules are actually often cheaper than LoRa-compatible ones) and deploy a set of end nodes, the only subject that can create and operate a SigFox network is SigFox itself, who sells access to the network and relevant additional functionalities on 'as-a-service' basis.

Deciding which model is better (between 'open' networks like LoRa and SigFox) is very much dependent on the specific solution that is being implemented: from one side with SigFox all the burden of deploying, maintaining and supporting the network is moved to a third party. This sort of outsourcing can become a competitive advantage when use cases involving very large areas to be covered are considered and, since it not possible to simply extend autonomously the SigFox network, monitoring and controlling on the actual coverage on the targeted installation area is mandatory. On the other side, other solutions like LoRa give the possibility of creating private, personally owned LPWAN networks, without having to share resources with anybody else (and without the need to pay recurring fees to a third party). Also, business continuity should be taken into consideration since, in case of severe problems to the mother company, availability of existing SigFox networks could be at stake.

From a technical standpoint, SigFox uses binary phase shift keying modulation (BPSK) in an ultra narrowband located within the ISM frequencies ranges. The utilization of an ultra narrowband allows for a simplified design for the radio components, and in particular the antenna, reducing the noise levels and the power consumptions. The downside is the available data rates: uplink data rate is limited to 100 bps with an additional limitation (due to the licensing model of the ISM band) of no more than 140 uplink and 4 downlink messages per day; the difference between uplink and downlink means that acknowledgement of the messages, as well as other functions like remote firmware update of the end nodes, is not possible. Coverage is, on the other side, a strong benefit of the radio technology used by SigFox: several studies have shown coverages ranging from 10 km in urban environments to 50 km for rural areas.

SigFox networks are only based on star topology: an end node can be connected to more than one base unit, but no mesh or combined topologies are supported. SigFox has recently enhanced its offering with a set of additional services, including data processing, big data analytics and devices geo-location. Just like what LoRa claims to do, also in this case, the position of the devices is calculated using a probabilistic calculation of the device position done using radio parameters. SigFox claims accuracy of this solution to range between 1 and 10 km depending on the base station density where the device is located. Whether this is enough or not will, of course, depend on the specific application being deployed.

### 1.2.2.2    Cellular technologies

Cellular communications have been growing at very high rates from the beginning of the 2000s; voice and short messaging services have progressively left to faster and faster data connections, the flag of being the driving force for the on-going expansions of the cellular networks.

Nowadays, three main families of cellular technologies are available:

- Global system for mobile communications (GSM): the first real widespread technology for cellular communications, mainly focussed on voice messaging, also known as 2G;
- Universal mobile telecommunications system (UMTS): still very focussed on voice and traditional messaging (through SMS), it introduced significant improvements to cellular data utilization in terms of available bandwidth and connection reliability, also known as 3G and
- Long-term evolution (LTE): an IP-only based cellular platform allowing even higher data rates and faster connection times, also knows as 4G.

Because of their nature, though, cellular networks technologies we are used to    AQ1
such as GSM, UMTS and LTE are not a particularly well fit for IoT applications, power requirements and complexity of a GSM modem, for instance, are not compatible with scenarios where constrained devices would be used, and the most important characteristics of a cellular data connections (e.g., support for high-speed mobility and high data rates) lose most of their value in several IoT scenarios. Even if cellular networks have been used in several M2M applications (for instance using SMS for data transfers, instead of proper data connections), many of the most important IoT requirements were left uncovered, leaving space, as we have seen, to other LPWAN technologies to emerge and take the field. For these reasons, in 2016 the third generation partnership project (3GPP, i.e., the organization that generates and maintains standards for cellular networks) released a series of specifications for IoT-specific standards, with the clear goal of making possible for the MNOs to be competitive in the IoT arena. LTE-mobile (LTE-M) and narrowband IoT (Nb-IoT) are the two IoT-specific standards, based on LTE technology, released by 3GPP in 2016. They have been both specifically designed for IoT applications, but with different specifications in terms of available data rates, support for mobility and link coverage.

The biggest competitive advantage for both LTE-M and Nb-IoT relies on the fact that both standards have been designed to be implemented on existing LTE networks with very limited or no changes to the existing network hardware. This feature will allow MNOs to upgrade their networks introducing LTE-M and Nb-IoT, without having the need of physically changing any of their equipment, thus reducing the implementation costs, making at the same time advantage of the existing installations and network coverage.

At the moment of writing, several MNOs have deployed pilot or commercial LTE-M and Nb-IoT networks, but for Nb-IoT there are very few modules commercially available that can use them, and the ones that are available are quite expensive, especially when compared to other LPWAN, noncellular technologies (100 US$ for an Nb-IoT module against 5–15 US$ for a SigFox or LoRa-compatible module).

## LTE-M

LTE-M has been the first IoT-oriented standard released by 3GPP and it is closer to LTE in terms of functionalities, hence representing a sort of the first step into the IoT domain for LTE-based cellular networks. It supports extended battery life through the implementation of specific functionalities like the power saving mode (PSM) and the extended discontinuous reception (eDRX), introduced in order to preserve battery life. On paper, all power saving functionalities set to the maximum could allow to reach almost 10 years. LTE-M offers higher data rates (up to 1 Mbps) that could support also voice-over-IP applications, making it different from the other LPWAN technologies that we have seen until now. On top of that, support for mobility (i.e., seamless handover of communication between nodes when the device is moving) is supported. While LTE-M was originally intended to be a sort of temporary solution to enable IoT on cellular networks, its capabilities allow for more flexibility than Nb-IoT and some mobile operators are actually considering LTE-M as an interesting solution for a wide range of applications in a future-proof way.

## Nb-IoT

Nb-IoT is the narrowband standard specifically released by 3GPP to support IoT applications. It reuses a good number of LTE characteristics, removing at the same

AQ2

time other functionalities like Inter-RAT handover (i.e., the possibility for an LTE device to jump on GSM or UMTS networks) in order to allow for simplified, and therefore cheaper, device design. Nb-IoT uses LTE licensed frequencies and can be deployed as stand-alone, in-band or in guard-band, making possible to the MNOs to adjust frequencies utilization based on the specific conditions of the area where the network is deployed. The design goals for the radio interface of Nb-IoT are indoor coverage, high number of connected devices and large coverage radius per eNodeB (i.e., the transmitting base station for an LTE network). Being based on LTE, radio access is more complex than what is required for other technologies like SigFox and LoRa; complexity in this case also means more expensive hardware, but at the same time better performances in terms of data rates. Nb-IoT can, in fact, reach 250 kbps which is significantly higher than what is offered by many other competitive technologies. Nb-IoT embeds other functionalities that are designed to minimize access latencies to network and to speed up connectivity set-ups, simplifying some of the procedures typically required by cellular networks to set up a data link (specifically attach and pdp context activations).

Being based on an LTE design, Nb-IoT requires a subscriber identity module card (SIM) to be included in the device. While standards and functionalities to implement software-based SIMs are available, at the moment of writing SIMs are mainly physical and this is a factor that could introduce additional complications in an IoT scenario (Table 1.2).

## 5G

With massive tenders going on around the world to assign the different frequencies bands to the MNOs, it is worth spending a few words on the next evolution of cellular technology, named fifth generation (5G) in order to distinguish it from 4G. The main focus of 5G will be on even higher data rates and lower latencies than what is currently offered by 4G networks. 5G main applications will be therefore in areas where either very high throughputs are required or very reliable and fast connections are needed for security reasons, like for autonomous driving and vehicular traffic management. 5G is therefore born having IoT in mind, but it will take a significant amount of time before large commercial applications may be available using this technology.

Table 1.2    *A summary of some of the characteristics typically considered when comparing long-range communication protocols. For coverage, typical utilization scenarios have been used to make a qualitative evaluation. Notice, however, that for some technologies (e.g., Nb-IoT) limited information on real-life applications are available at the time of writing*

| Protocol | Packet size | Data rates | Coverage | Topology | Notes |
|---|---|---|---|---|---|
| SigFox | 12 bytes defined by user | 100 bps[1] | 3–10 km (urban); 30–50 km (rural) | Star | Networks managed by SigFox |
| LoRa/ LoRaWAN | – | 0.25–50 Kbps | 2–5 km (urban); 15 km (rural) | Star of star | Network can be created and maintained privately |
| LTE-M | Defined by user | 1 Mbps | 2.5–5 km | Cellular | Network access regulated by contracts with MNOs |
| Nb-IoT | Defined by user | 250 Kbps | 15–30 km | Cellular | Network access regulated by contracts with MNOs |

[1]With limitations on the number of messages/day.

## 1.3   Data protocols

Besides the radio and networking protocols that allow the interconnection and logical organization of the devices in the networks, the protocols that stay on top of the protocol stack are extremely important as well, since, at the end of the day, it will how they access the transmission layer to determine the actual dynamic characteristics of the IoT application. Similarly to what we have seen until, there is no simple answer, nor one catch-all solution that fits for all applications, but rather different options that may fit better one use case instead of another.

### 1.3.1   Hypertext transfer protocol

Hypertext transfer protocol (HTTP) is one of the most famous protocols in the world, well known also to the ordinary persons, being the most visible protocol used to browse resources on the web. The first version of HTTP was released 0.9, dated 1991 and even at this early stage the foundation principles of that made of HTTP such a huge success were already present: simplicity and the response-request structure.

HTTP is, in fact, based on a client–server paradigm, where a client (or more properly, a user agent) wants to access to resources located on a remote machine (the server). In order to access those resources, the client must formulate a request to the server, specifying which resource it wants to access using a resource identifier (called uniform resource identifier, or URI in short). The server must respond to this request specifying if the request has been successful or not and, in case of success, providing the resource itself.

HTTP specifies few methods that can be used by the client to interact with the server: these methods are used to instruct the server on the operation that the client wants to accomplish with the specified resource (e.g., retrieve an HTML page). Resources can be static, that is, present in the server in a permanent (or semi-permanent) way, or dynamic, that is generated dynamically following an explicit request from the client. In this way, it is possible for the server to generate user or session-specific contents that can, for instance, embed data coming as a result of a query from a database. Another important aspect of HTTP is the fact that it is a stateless protocol: the server does not store information on the current session with a specific user agent and it is for the latter to keep track of the history of its interactions with the server (cookies have been introduced for this purpose in more recent releases of the protocol).

HTTP evolved from version 0.9 to 1.1, released in its first version in 1997, which has been (and still is) the standard used for the world wide web (WWW); notwithstanding the various improvements that have been introduced over the years, the key principles outlined above have remained more or less unchanged. Using HTTP for IoT applications is possible (indeed, HTTP is already used by several IoT applications) but the utilization of this protocol in this domain comes with a number of concerns:

- the client–server architecture implemented in HTTP does not support server push – there is no support for a spontaneous data transfer from the server to the client;
- headers (i.e., the commands and the responses used by the two parties to communicate with each other) are transferred in plain text and are quite verbose. While this can be an advantage for a human being looking at an HTTP interaction, for constrained devices utilization of this type of headers can easily cost valuable resources both in terms of memory and bandwidth utilization;
- HTTP only allows one request at the time to be processed, generating unnecessary delays when processing multiple requests at the same time and
- HTTP is designed to run over TCP, which can be a resource-intensive protocol for constrained devices (and may introduce unnecessary features).

In order to overcome these limitations, HTTP 2.0 (in short HTTP/2) has been designed. The new version, introduced in 2015 and already supported by the most important web browsers, overcome almost all limitations listed above, with the only exception of the forced utilization of TCP, which remains the network protocol of choice also for HTTP/2. In particular, release 2 of HTTP introduces server push (essential for publish-subscribe architecture, see MQTT) and headers compressions in order to save bandwidth in constrained devices. Encryption is not considered as mandatory in the standard but several implementations (including all the most popular web browsers) will not support HTTP/2 without an encrypted layer like TLS.

Using HTTP/2 for IoT has interesting benefits in the form of using a widely adopted protocol that is supported by a large number of platforms and software systems, at the same time for use cases where bandwidth and processing power are scarce other options (like CoAP or MQTT) may be more efficient.

## 1.3.2   Constrained application protocol

As mentioned in the previous paragraph, HTTP/2 alleviate most of the major drawbacks that prevented HTTP/1.x from being successfully used in IoT applications. Nevertheless, it still requires TCP as its transport layer, despite admitting that the latter was not specifically optimized neither for IoT applications nor for constrained devices. In order to overcome these limitations, the constrained application protocol (CoAP) has been designed. Officially defined in RFC 7252, CoAP has been designed from the very beginning to be

- compatible (i.e., operational on) with constrained devices, typically 8-bit microcontrollers with 10 KB of RAM and 100 KB of code;
- based on UDP (much lighter in terms of resources utilization than TCP). CoAP is based on the REST paradigm (REpresentational State Transfer), an architectural style based on the typical client–server paradigm of HTTP extended to web applications. In simpler terms, CoAP and HTTP share the same request-response messaging scheme, making interconnection between HTTP-based systems (such as web servers or web applications) and CoAP domains a relatively easy task. The basic scheme of operation of CoAP can be divided into two logical layers (even if the protocol layer remains only one);

- Messaging: this layer provides the tools for message transfer between the different nodes. Messages can be transferred in a reliable (i.e., with an acknowledge) or unreliable way – it is up to the application to decide which option to take (differently from TCP which 'forces' acknowledgement of every message). The availability of reliable message dispatch overcomes the lack of the error and flow control in UDP (present instead in TCP) and
- Request-response: this layer provides a way for the parties to publish (i.e., post) a value to the server, to obtain from the server information (identified by a URI) and to subscribe to a specific information, implementing a server-push mechanism.

An interesting functionality included in CoAP is the inclusion of discovery functions that can be used by a node to discover other nodes currently belonging to the same CoAP domain. This functionality can be useful when building dynamic networks of devices over, for instance, unstable or not continuously available connections.

It is interesting to note that even if CoAP is designed to run on UDP, it can actually be implemented on any network protocol since the CoAP messaging layer implements the required functions of flow and error control; this way, it is possible to use SMS (for instance) to relay CoAP messages over a cellular network saving the client from the additional burden of handling UDP.

Security is not defined in the CoAP specifications and, when implemented over UDP, security functions are implemented through the utilization of the datagram transport layer security (DTLS) – this may, however, represent a drawback since implementing DTLS in constrained devices may hinder some of the benefits portrayed by CoAP.

### 1.3.3   *Message queuing telemetry transport and MQTT-SN*

Message queuing telemetry transport (MQTT) is a lightweight message-based communication protocol based on a publish-subscribe paradigm and especially developed having in mind constrained devices. MQTT, in fact, has been designed to use as little bandwidth as possible: specifically, the primary requirement for this protocol was to use less bandwidth than that required to carry out the same actions using HTTP or similar protocols. Tests in this direction have been carried out, highlighting that utilization of MQTT results in more messages being transferred (in a given period of time) with lower power consumption when compared to HTTP/1.x. When it comes to HTTP/2, the header compression mechanism introduced in that release of the protocol actually limits the bandwidth used by the nodes, but increases the computational requirements, leaving MQTT to remain the most sensible choice for constrained devices.

As mentioned above, MQTT is based on a publish-subscribe paradigm; this means that the available information is classified in topics. For ease of information access, each topic may be positioned inside a hierarchy (similarly to what happens in a normal file system with folders). Distribution of the information is not therefore based on point-to-point (or even point-to-multiple points) model, but rather on the contents of the information itself.

In a basic MQTT network, there are two different types of elements: nodes and brokers. Using a typical client–server metaphor, the broker acts as a central server for all nodes (acting as clients) and all nodes communicate (both in read and write modes) only with the broker. It is the broker's responsibility to distribute the information accordingly to the 'interests' of the nodes.

A node can, in fact, be both an information producer and a consumer, even at the same time: nodes can subscribe to specific topics of interest, in order to make sure that the broker will send them all the messages pertaining to those topics. Similarly, nodes can create contents by publishing information to specific topics. The broker does not carry out any processing on the transmitted message, as it simply relays the payload to all nodes that previously subscribed to the topic.

An interesting advantage in respect of more generic protocols, like HTTP/2, is the inclusion of specific functionalities for message and telemetry transfer over troubled connections, like the 'last will and testament' (LWT) functionality, which allows to specify a message that the broker has to broadcast to all interested parties when a client is disgracefully disconnected from the network. The LWT allows therefore to implement fall-back strategies in specific circumstances.

MQTT relies on TCP as transport protocol, and as such requires a TCP layer in the lower protocol stack to operate – for applications where the implementation of a TCP stack can be problematic, MQTT-SN (MQTT for sensor networks) has been designed, introducing solutions to allow the application protocol to be agnostic from the lower levels.

Even if MQTT is widely used in a number of applications, one of its main limitations is about the lack of security measures specified into the standard, leaving the definition and the implementation of security measures to the specific implementation. While this leaves freedom and helps to implement the protocol on particularly constrained devices, it is important to note at the same time that security is becoming more and more an issue for IoT applications and as such inclusion of security mechanisms such as TLS is as a matter of fact almost mandatory.

## 1.3.4    Extensible messaging and presence protocol

Extensible messaging and presence protocol (XMPP) was born as an open-source effort to standardize interoperability between different systems using the eXtensible Markup Language (XML), providing a way to implement messaging and presence system using XML files. XMPP has been later standardized by IETF which has issued several RFCs that define various aspects of the protocol, including specific implementation notes for IoT applications.

Each node belonging to an XMPP network is identified by a unique identifier, composed by a user id and the identification of the XMPP domain. Each user can also have multiple devices connected to the same account with different priorities, making possible selective dispatch of messages to the different devices in a specific order (e.g., the 'main' device can receive all messages while others can receive only specific ones).

The key-characteristic of XMPP is, as already mentioned, the adoption of XML files to execute all operations, from nodes binding to information request and subscription. Requesting information from a node is carried out by sending an XML file containing the requester node identifier, the recipient identifier and an instruction to read a certain information element. The recipient node will then answer with another XML file with the response, being that the information itself or a message describing the status of the node and why the information could not be provided.

The key advantage of using XML is the possibility of implementing a much more complex hierarchy of data and queries with respect to what is allowed by other protocols using 'flat' data representations (e.g., MQTT). This makes possible, for instance, to specify different data types (e.g., integer, Boolean, string, etc.), to simultaneously read (or write) multiple information with a single operation, and to provide additional information about the data being read or written (e.g., if the data are an actual read from a sensor, an averaged value, an estimate, and so on).

This added level of flexibility brings an inherent increased level of complexity when comparing XMPP to other message-based protocols like MQTT: each node must be able to properly read, parse and manage XML files, and possibly manage the implemented data hierarchies. Even if specific measures have been defined for implementing XMPP in IoT scenarios, the added requirements in terms of bandwidth and computational power need to be taken in account when designing the application, especially if constrained devices are being used.

XMPP per se is defined as a very generic and extensible protocol and can be therefore adapted to a number of applications. Recommendation XEP-0232 explicitly targets utilization of XMPP for IoT and covers a number of areas such as definition of XML tags that preserve readability but reduce the overhead required to transfer XML files (typically larger in size than binary files). Even though XMPP foundation claims that XML parsers are typically as efficient as JSON parsers (or even more efficient), the very descriptive nature of XML requires a certain overhead for transmitting data.

Considering all of the above, many authors concur in considering XMPP an interesting alternative when scalability, data complexity and interoperability are key-requirements and where computational capabilities of the nodes being used are not a major design concern; for other applications, more simple and straight-to-the-point protocols (like MQTT or CoAP) may represent a better alternative (Table 1.3).

*Table 1.3    A summary of the main features of each data protocol analysed*
*in Sections 1.3.1–1.3.3, from an IoT application standpoint*

| Protocol | Type | Pros | Cons |
|---|---|---|---|
| HTTP/1.1 | Generic | Widely available<br>Easy interoperability with web applications | Uses TCP<br>Bandwidth overhead because of headers |
| HTTP/2 | Generic | More efficient than HTTP/1.x<br>Maintains interoperability with web applications | Uses TCP<br>Headers compression requires computational power |
| CoAP | Messaging | Compatible with constrained devices<br>Can use UDP or other lower transmission protocols<br>Specific for IoT | Security not included in the standard |
| MQTT | Messaging | Compatible with constrained devices<br>Specific for IoT | Security not included in the standard<br>Several different implementations may cause interoperability issues<br>Uses TCP |
| MQTT-SN | Messaging | Compatible with constrained devices<br>Specific for IoT<br>Can use UDP or other transport protocols | Security not included in the standard<br>Several different implementations may cause interoperability issues |
| XMPP | Generic | Extensible<br>XML allows for more complex data management | Overhead introduced by XML<br>More complex w.r.t. other protocols |

## 1.4    IoT platforms

As lots of IoT platforms are currently available in the market, it is difficult to decide which one to go for. Almost all of them are capable of saving sensor data into the cloud and performing basic functions like predictive analytics, machine learning, visualization, etc., on top of it. However, having a deep dive into these platform solutions one will understand the core difference between the well-established and the basic products available in the market. To be precise, there are certain technical as well as commercial aspects which define the maturity of an IoT platform and must be taken into consideration while selecting an appropriate solution for a specific IoT use case.

### 1.4.1    Technical aspects

#### 1.4.1.1    Services offered

Almost all platforms offer additional services on top of the basic data acquisition functionalities. These services typically include analytics functionalities as well as standard plug-ins or applications such as location tracking, error notification, real-time monitoring and so on. Considering therefore the set of the potential vertical functionalities, all platforms try to offer the same menu of main options. What differentiates a mature platform solution to others is the actual type of analytics (real-time, batch, predictive, etc.) and the number and diversity of the plug-ins already available and embedded into the platform. Some well-established platforms, for instance, offer industry-specific (aerospace, shipping, manufacturing) plug-ins in their portfolio that simplify the whole process of setting up the system.

Another important feature of an IoT platform is its openness (i.e., being open source or not) to allow the users to launch their own personalized plug-ins or APIs. Not every platform is based on an open-source environment and having to deal with closed, protected platforms may represent a significant issue when trying to make the platform evolving with the real-life application behind it. The ones which allow the users to launch their own APIs offer different Software Development Kits (SDK) to program these APIs. Certain mature platform solutions have SDKs which includes open-source libraries in Java, C, Python, etc., giving developers options to program their own solution using a platform of their choice [11].

#### 1.4.1.2    Supported data protocols

IoT applications works on various data protocols depending upon the type (continuous, sensor data, audio/video, etc.) of data to be monitored and the required degree of security (TLS/DTLS) that needs to be achieved. As we have seen in Sections 1.3, among the most common data protocol for IoT cases are MQTT, HTTP and CoAP. In contrast to HTTP, MQTT and CoAP are protocols best-suited for a constrained environment where the bandwidth and sensor battery are at a premium. MQTT is used extensively for event-based monitoring and CoAP is a preferred protocol for continuous data transmission [11].

Not every IoT platform supports all of these protocols, which also means that the one covering all of them may have a significant competitive advantage in situations where flexibility and systems' interoperability is a key requirement.

#### 1.4.1.3    Security

Security is a factor which cannot be ignored or underestimated. Since IoT is all about data and data management, the platform, as well as the underlying infrastructure, must ensure data integrity and safety. At the same time, no loose points should exist in the IoT infrastructure which can later be used as a backdoor entry for hackers which may then not only try to steal the data but can also manipulate the system. The balance between data integrity and data protection is defined by the specific application, but the best approach is to choose a solution that enables both aspects at the same time.

IoT security should therefore be implemented at every level in the IoT stack starting from the sensor-gateway and up to the cloud-data centre. A good platform solution uses unique identity-based authentication methods such as X.509 certificates at sensor/device level and offers encrypted (TLS/SSL) data transfer within the IoT network. Constrained devices may use other dedicated solutions that preserve the required data security functionalities while at the same time avoiding the overhead caused by some of the already mentioned security solutions (e.g., TLS/SSL). Some IoT platform solutions offer special authentication methods on the top of what is typically considered to be a standard set of features and can offer compliance with security management systems such as ISO 27001.

### 1.4.1.4   Device management

Device management is an important feature of an IoT platform that, interestingly, is not offered by some of the solution providers. It is a feature which maintains a list of devices connected to the IoT network and keeps track of their operational status. Moreover, this feature often includes remote devices provisioning and configuration management, allowing remote configuration of a device, firmware updates and faults management (detection and first-level troubleshooting) in the device.

Another important aspect of the IoT platform which is directly associated with Device Management is its degree of scalability. Adding more devices to the network increases its complexity and demands scalable storage along with load balancers (to distribute data load over several servers). Maintaining these servers, load balancers and subsequent data nodes is not easy. While Device Management is certainly a functionality that needs to be taken into account for large and complex systems, for smaller ones it may not be as essential. For this reason and its inherent implementation complexity some IoT platform solution providers do not offer device management capability [11].

### 1.4.1.5   Data storage

Data storage is an aspect obviously related to data security but that extends also to other specific (and very important) aspects that need to be considered, such as data availability, backup and easy recovery capabilities. All these capabilities derive directly from the IT infrastructure being used, that may include concentrated or distributed storage systems with different levels of resilience implementation. Backup solutions can also vary from simple backups (complete or incremental) to full disaster recovery solutions, with copies of data in multiple physical sites and complete solutions in place for fast and error-free system-wide recovery. Cloud platforms further extend this number of available solutions, including also the possibility of integrating cloud platforms with on-premises storage (hybrid cloud). The hybrid cloud option could allow the user to process and store business sensitive information locally and keep the remaining mass data inside the IoT platform cloud, obtaining at the same time interesting results both in the data security and data reliability domains.

In order to minimize the data load, certain platform supports decentralized, offline data computation, better known as edge computing (see Sec. 1.5). This feature will play a very important role in the future when IoT projects will get bigger and more complex: in this scenario it will be important for a redistribution of the computational and storage demands in order to ensure continuous scalability of the platforms. Edge computing not only allows for lower operation costs by reducing the data load being pushed to the cloud but also enhances the application performance by achieving lower latency levels.

## 1.4.2    Commercial and business aspect

### 1.4.2.1    Size of the company

Thumb rule says, the bigger the brand, the firmer they have to stand for their reputation. Which means that they strive to provide a quality solution. The same applies to the support and maintenance services, that are now more and more considered as an integral part of the features offered by a solution. Service level agreements (SLAs) offering global, 24/7, multilingual tech support are something which really differentiates big, well-established companies from newcomers or start-ups and while additional services come with an extra price tag, for complex and large systems, a well-performing support and maintenance system can make the difference between a usable and an unusable system.

### 1.4.2.2    Data centre

Certain big IoT platform providers manage their own set of data centres worldwide. This is particularly relevant if you are considering *'one-stop'* solution (e.g., platform + cloud) rather than dealing with two or more parties in order to host your application. Dealing with one party is of course commercially beneficial and easier than negotiating deals with two or more parties.

The idea behind one-stop platform plus data centre concept is also important if you are looking for a long-term, stress-free and reliable solution. The platform providers who host their solutions on a rented cloud/data centre may change their opinion with regards to the choice of data centre provider later thus resulting in troublesome data migration.

A third important aspect of the data centre is the availability of data worldwide with as low latency as possible. Some of the bigger IoT platform solution providers have data centres installed in every continent, with automated data replication functionalities included in the solution, thus addressing this issue [12].

### 1.4.2.3    IoT ecosystem

IoT ecosystem is currently one of the most important selection criteria for a platform solution. An ecosystem in IoT terms is actually the capability of an IoT platform to bring the partners of the value chain (e.g., stakeholders, market players, etc.) together or to integrate their solution in the platform in order to broaden the overall digital product portfolio. This is especially important if a company would like to integrate its solution and the solutions of its suppliers and service providers with customers at one place. Here again, the bigger the brand of an IoT provider, the larger its business network thus offering bigger ecosystems.

Another important aspect of IoT ecosystem is the interoperability. A good IoT platform solution should support integration with open-source ecosystems and should provide an interface for third-party services or applications. This aspect becomes particularly important in cases when the platform solution is needed to integrate with the existing ERP or other already existing database management systems [12].

#### 1.4.2.4    Pricing model

The most common pricing model offered by platform providers is subscription based on a *pay-as-you-go* scheme (these platforms are also often labelled as *PaaS*, or Platforms as a Service). This model can be effective and attractive for small systems or for portions of larger solutions (imagine for instance a section of a monitoring solution that needs to use satellite links for a part of the data transfer), but of course does not scale particularly well when the size of the system increases, since the costs will easily and quickly become non-negligible. Another option is represented by a subscription model (i.e., with a fixed monthly fee) that offers an easy way for managing the cost and provides effective revenue projection. To summarize, the pay-as-you-go model is good for users who are well aware of the amount of data load they want to handle for a long period of time, whereas the subscription model is a better alternative for highly unpredictable, complex and fast-growing IoT use cases.

Some platform solution providers offer both pricing models in order to give the customer a choice of matching cost structure based on his individual use case. Flexible pricing models are more typical of newcomers or start-ups, that are using flexibility as a competitive advantage to gain access to the market.

### 1.4.3    Commercial or open-source

As already mentioned, many of the commercial IoT platforms are based on a *pay-as-you-go* scheme, based on parameters that scale with the amount of traffic (e.g., number of messages, number of connected devices, etc.). This of course means that the price of the solution will scale (linearly or not depending by the specific business model in place) with the size of the system hosted by the platform. As an alternative, there are several (literally hundreds!) of IoT open-source platforms that could represent a viable alternative to commercial solutions.

#### 1.4.3.1    Evaluation criteria

However, there are at least three factors that need to be taken into account when making a decision about the platform to be used:

• technical aspects;
• projected growth of the system and
• commercial considerations.

The technical aspects include the basic requirements that the platform needs to fulfil, including, but not limited to

• reliability: how reliable is the platform? how large is the installed base?
• scalability: how easy is it to scale up (or down) the platform?
• protocols: which are the supported protocols? How easy is it to add a new protocol to the platform?
• support: what are the available SLAs? Is the support localized (i.e., available in your local language)?

- supported cloud platforms: which are the cloud platforms/interfaces compatible with the platform? In other words, does the choice of the platform 'lock' you to a specific cloud solution?
- security: what are the available security protocols/mechanisms? and
- ease of use: how easy is it to manage the platform and the devices? Does the platform include a provisioning or O&M section to allow for simplified devices management?

It is clear that the list could be further expanded, but the above points are most probably the most important ones to take in consideration. In particular, reliability is a point that is typically in favour of commercial platforms since large IoT providers invest a significant amount of money in their data centres, offering therefore availability and reliability figures that can be difficult to be matched by open-source alternatives. On the other hand, choosing a commercial platform may complicate future migrations to other solutions, since data export and migration functionalities may be limited (or completely absent).

The projected growth of the system has an impact, obviously, on the already mentioned scalability factor, but also on the other supported protocols. To make an example, for a small system, the lack of a specific communication protocol or the lack of a device management function may not be a problem, but the difficulties and the overhead caused by these missing functionalities may explode at a later stage when the size and the complexity of the system will increase. It is therefore important to have (even a minimal) vision of the potential growth directions of the system, both in terms of size (e.g., number of devices) and complexity (e.g., number of protocols to be supported).

Last but certainly not least, the commercial aspects may play an important role in the choice. We have already analysed the pricing model for the commercial platform; for open-source, there are essentially two factors to be taken into account:

- self-hosted or server-less and
- pricing model.

The first point refers to how the platform will be hosted: in a self-hosted the owner will have to sustain the initial costs for the hardware and storage, connectivity fees (i.e., Internet access) and all the related operational costs (e.g., maintenance, power, air conditioning, etc.). Considering that the upfront costs associated with the acquisition and set-up of a hardware platform supporting high-availability and AQ3 redundant data storage can be significant unless there are specific requirements (e.g., security or data confidentiality) that prevent the utilization of the cloud, a server-less implementation becomes almost a mandatory solution. In this case, the servers are hosted by a cloud provider; also in this case we typically have the choice between two business models:

- *pay-as-you-go*: in this case the price is calculated using a fixed based on the size of the storage, the number of CPUs, RAM, etc. plus a variable amount calculated on the number of data being transferred and
- fixed subscription: in this case, the fee covers the virtual hardware configuration and a *flat rate* of data per a certain period of time (e.g., a month).

There may be additional costs to be added to the basic subscription, like enhanced availability, extended support SLAs, etc. In any event, it is clear that, once again, having a view of the initial size and projected growth of the platform becomes a fundamental aspect, since it will not initially allow to choose between the two models but will help to predict about the future costs and possible migration directions. The *pay-as-you-go* model will in fact scale with the system's size and there will be a cut-off point when the fixed subscription model will become more financially attractive. Predicting when this transition may happen can become an important factor to ensure growth sustainability of the platform.

### 1.4.3.2    Open-source platforms

As already mentioned above, there are literally hundreds of open-source IoT platforms available, with a business model that is typically based on support and professional services (e.g., training and consultancy) fees. Choosing a specific platform may therefore be a complicated task, made even more complex by the fact that not all the platforms have comparable attributes in terms of the requirements listed in the previous paragraph. We will nevertheless mention here a limited set of the most common open-source solutions, with the goal of presenting the ones that bring particular functionalities in one or more areas.

*Kaa*: Kaa is an open-source platform that supports a wide range of endpoint systems (including Raspberry Pi, Intel Edison and ESP8266, but not Arduino family) [13]. It is possible to extend the support to the platform by using SDK and porting it to the target platform in case this is not included in the list of systems supported natively. Kaa uses Apache ZooKeeper to create and coordinate Kaa clusters, and SQL and NoSQL databases for configuration and data management. The implementation of a cluster structure allows the creation of high-availability (HA) solutions, that have to include also the database instances. The entire code is available for download on a git repository and there are active communities supporting and using Kaa, including a dedicated forum on Stack Overflow. In terms of supported protocols, Since Kaa is intended to be a middleware component, analytics components are not included in the distribution but can be easily connected to the database using, for instance, Apache Zeppelin. In terms of supported protocols, communication between endpoints and the Kaa cluster is managed through the Kaa SDK, but the platform supports MQTT, CoAP, XMPP and HTTP (all over TCP).

*SiteWhere*: SiteWhere uses a number of well-established technologies to provide a platform that facilitates devices management and different aspects of data generation, storage and distribution [14]. An interesting aspect of SiteWhere is the utilization of an In Memory Data Grid (IMDG) solution to provide subscription-based access in real-time to the data feed being generated by the devices. Permanent storage is achieved thanks to the utilization of nonrelational (e.g., MongoDB, Apache HBase) and/or time-series database components, that can be installed, together with the key components of SiteWhere, in a cluster configuration in order to implement HA. Communication between the end device is carried out through two pipelines (one for inbound messages and data and one for outbound commands to the devices), and the list of supported protocols include MQTT, AMQP, Stomp and WebSockets. SiteWhere does not include analytics components but rather relies on external ones that can be connected using MuleSoft AnyPoint connectors.

*Eclipse IoT projects*: Eclipse is a very well-known name among developers, particularly for being the house of one of the most used IDEs for Java. Within the Eclipse foundation several other projects are hosted and developed, including a set of IoT-related projects supported by a vibrant community. The home page of the Eclipse IoT project includes in fact several projects, all open-source, that cover almost all aspects of the lower layers of a typical IoT project, from devices and protocols to gateways [15]. Worth mentioning, among all the others, are as follows:

- Paho: an implementation of MQTT and MQTT-SN in several languages, including Java, C, C++ and C# [16];
- Californium: an implementation of the CoAP protocol in Java for nonconstrained devices [17];
- Hawkbit: a tool for roll-out of software updates to remote devices [18];
- Kura: a container for M2M applications running in service gateways – includes a native MQTT-based messaging solution to allow communication with devices and remote management [19] and
- Agail/Agile: a modular hardware and software gateway for the IoT, with support for protocol interoperability, device and data management, IoT apps execution and external Cloud communication, featuring diverse pilot activities, Open Calls & Community building [20,21].

While the Eclipse IoT initiative does not have a self-contained platform project, it is possible to use one of the existing projects and build on top of that. A key advantage of such a solution would clearly be the very large community of Eclipse developers around the world that provide through different means support and examples.

*ThingSpeak*: ThingSpeak is the 'open IoT platform with MATLAB analytics' [22]. The platform includes support for several endpoint devices, including Arduino, ESP8266 and Raspberry, offering REST API for communication with external additional components. On top of that, it offers the possibility of analysing the data coming from the devices using the MATLAB analysis application, one of the components of the MATLAB suite particularly suited for analysis on mathematical and statistical data. Interestingly enough, you do not need to own a MATLAB licence to use its analytics module in ThingSpeak. ThingSpeak is actually not a completely open-source system since the code for the server and analytics is not available; there are instead git repositories for the communication libraries to be used in the remote devices. Another interesting functionality offered by ThingSpeak is the possibility of triggering a reaction (e.g., sending a Tweet) when specific conditions arise. This functionality is implemented through a rule engine that processes the data coming from the device and, when the conditions specified in the rules are met, triggers the programmed reaction.

*DeviceHive*: Distributed under the Apache 2.0 licence, DeviceHive comes in a ready-to-be-deployed docker with support for virtualization and provides a 'digital playground' for experimentation [23]. All code is available from a git repository, including device libraries that allow connection with the server using REST API, WebSockets or MQTT. Tutorials and documentation are available on the website, explaining also how to connect the platform with external modules to implement batch analytics and machine learning on top of the data coming from the devices. Though DeviceHive is not natively provided with an analytics solution, dataset interface for utilization with Grafana (an open-source software for time-series analytics) is provided (Table 1.4).

*Table 1.4   A summary of the most common open-source solutions, together with some of their key features*

| Platform name | Device management | Supported devices | Protocols | Analytics | Database |
|---|---|---|---|---|---|
| Kaa | Yes | Raspberry, Intel Edison, 8266, … | MQTT, CoAP, XMPP and HTTP | Through external tools | SQL and NoSQL |
| SiteWhere | Yes | Unclear | MQTT, AMQP, Stomp and WebSocket | Through external tools | IMDG and NoSQL |
| Eclipse | Yes | Several | MQTT, CoAP | Through external tools | External |
| ThingSpeak | No | Arduino, Raspberry, ESP8266, … | REST and MQTT | MATLAB analytics | MySQL |
| DeviceHive | No | ESP8266 | REST and MQTT | Through external tools | PostegreSQL and SAP |

## 1.5   Future directions and challenges

Despite the huge research efforts invested so far within the IoT research field, there still persist several open issues that are under hot debate and actively investigated by the community, gathering diverse skills and competences from Computer Science, Electrical Engineering, Electronics and Physics, among many others. We heretofore list such challenges, along with a short summary of the current developmental status of each line in regards to the latest findings.

### 1.5.1   Fog/Edge/Cloud Computing

Given the *in crescendo* variety and complexity of tasks to be undertaken by IoT devices, there has prevailed a strong interest in the computational implications of the deployment of such tasks on different levels of the IoT architecture. This has been of utmost relevance when dealing with highly constrained use cases demanding particularly highly efficient communication protocols and/or processing functionalities. This can be exemplified by delay-sensitive communications in which, should the processing capabilities of IoT nodes be lower than required, the design criterion of the IoT architecture should prioritize the deployment of all resource-consuming processing tasks upstream in the cloud infrastructure, thus leaving IoT nodes mostly dedicated to monitoring, communication and/or actuating procedures. On the contrary, if IoT nodes are empowered with more processing capability, a fraction of the tasks could be performed as close to the IoT node as possible, giving rise to the so-called edge/fog computing paradigm [24]. This option should also be selected when the network traffic is a limited asset, due to either stringent radio propagation conditions of the communication channel (as in e.g., industrial M2M communications) or the aggregate amount of data captured by the IoT mesh. Furthermore, these computing alternatives find their rationale not only in exemplificative general use cases as the ones used above, but also in other casuistry such as applications dealing with time-sensitive data, private information that should be stored and kept as locally as possible, or federated computing and distributed inference/learning aiming at extracting knowledge from the collected information in a distributed manner [25].

All in all, several questions remain unanswered in this active field: to begin with, the capillarizations of processing functionalities to the edge of the IoT architecture comes along with a complex, decentralized management of possibly heterogeneous devices. This poses fundamental challenges regarding interoperability, programmability, service migration under mobile roaming, network virtualization and scalability, far beyond and more involved than the usual management aspects yielded by distributed device management (namely, updating, patch versioning, etc). Other challenges related to the implementation of fog computing in IoT environments include those related to data governance, data privacy, data marshalling and data delivery among nodes for distributed computation, and the consideration of economic criteria in the choice of one processing implementation or another under different cloud pricing schemes, among others [24].

### 1.5.2 Ultra-reliable low latency communications (URLLC) and tactile Internet

By these terms, closely related to each other, we refer to all communication scenarios demanding extremely low latencies in combination with high availability, reliability and security levels [26,27]. Tactile Internet extends further this envisaged scenario with haptic interaction and visual feedback so that objects can be manipulated remotely by the user, who feels the touch and strength of his/her action. These aspects have evolved from simple design optimization objectives to restrict themselves, especially in fields where the interactivity between the human user and the deployed devices is crucial for the success of the application itself, such as industrial robotics, transport systems, healthcare/surgery, education, augmented reality (AR), virtual reality (VR) and gaming, among others. In the case of IoT systems, the need for meeting ambitious goals in these processing domains impacts directly on the design on the whole architecture so as to improve the access delay and reliability in both directions of the communication link (uplink/downlink) and the provision of intelligence to predict contents, anticipate communication failures and in essence, improve the efficiency under which information is delivered to the network.

Unfortunately, even though the activity in this field is notable – particularly in what regards to provisioning such demands over future generations of cellular communications – URLLC and Tactile Internet are still far from maturity [28], with scarce practical prototypes using ad hoc underlying communication technologies. There is a long road ahead to be driven by the community towards exploring how IoT environments can support massive scenarios demanding URLLC/Tactile Internet, with a major focus placed on how to meet well-known operational thresholds imposed by the physics of the haptic interaction and visual feedback (e.g., the Motion-to-Photon 20 ms latency constraint in VR applications). Key elements to be explored will include the inclusion of new network elements for decentralized computation offloading (cloudlets), which in turn will unleash new issues such as online resource management, provisioning and task allocation [29].

### 1.5.3   Embeddable artificial intelligence

To provide machines with learning capabilities lies at the very core of the future of IoT environments, which has become a reality by virtue of the practical proliferation of IoT devices and the provision of a digital data substrate from which to learn, construct models and exploit the information acquired by IoT nodes. Although the history of Artificial Intelligence (AI) dates back to the early twentieth century, the evolution of learning algorithms has not grown in the last years towards more sophisticated models but has instead steered towards traditional models leveraging the availability of more powerful computing resources. This is, in fact, the case of the celebrated Deep Learning, which departs from traditional Artificial Neural Network models to implement highly dimensional neural nets capable, not only of capturing more complex data patterns from the collected data, but also to override any need for preprocessing and extracting predictors from the data themselves [30]. This has been accomplished partly by modified neuron models (allowing for fine-grained, recurrent information feedback). However, most of the success of Deep Learning models corresponds to the implementation of their learning algorithms efficiently on devices with unprecedented computation capabilities (e.g., Graphical Processing Units, GPUs) and unbounded consumption of other resources and assets such as energy or storage.

This is not the case for IoT devices, particularly when the processing architecture opts for deploying the learning algorithm as close to data as possible. In this case, not only the learning algorithm must run efficiently on the IoT node not to drain the battery of the device, but it must also allow for an incremental operation that dismisses any need for locally storing large amounts of data. From an application-agnostic point of view, the study of incremental online learning algorithms for stream processing has been at the forefront in the last decade, capitalizing on how models should be retrained when dealing with nonstationary data sources [31]. However, most studies do not consider how the learning algorithm should also take into account the remaining battery level of the device in which it is executed, so as to properly balance the between model performance and complexity. Some attempts have been reported in this regard, such as the Embedded Learning Library developed at Microsoft Research [32]. This recent big leap in embeddable AI suggests that we face an unexplored niche of research gravitating on the design of machine-learning algorithms tailored for resource-constrained systems as those in IoT environments.

In this same line of reasoning, a major momentum has been lately gained by the so-called Federated Learning concept coined by Google [33]: this technology enables to learn locally a prediction model and share it collaboratively among other devices in the network, while retaining all training data on the device in which the model was built. The motivation behind the creation of federated learning models concentrates again on eventual bandwidth and latency constraints in the network collecting the data, which pushes the processing paradigm to an edge computing architecture. When the computing power of a single node is limited and the network undergoes such bandwidth/latency constraints, there is no other way to proceed than to deploy intermediate processing platforms (e.g., *cloudlets*) or, instead, resort to collaborative model learning technologies such as the ones currently pursued in the literature. Albeit certainly promising, research lines around the so-called transfer learning and domain adaptation [34] – which could make the knowledge learned in one IoT node or group of nodes transferable to other IoT nodes – are still in their infancy and have not grown mature enough for their widespread adoption in severely constrained application domains.

In summary, research on AI model is opening a wide spectrum of possibilities and applications in IoT environments, which are closer to practicality than ever. Data availability is not a problem any longer for machine-learning models, hence processing constraints should grasp the attention of the research community in order to realize fully operational machine-learning functionalities in IoT applications.

### 1.5.4   Secure communications

It is clear that the proliferation of more and more connected IoT devices has ignited the number and capabilities of services and applications, but unfortunately also unchains an exponential increase of security threats and backdoors for malicious users to capture sensitive data and hack systems and processes monitored/actuated by IoT nodes. Such security holes, when discovered by cyber-criminals, can cause damage at a grand scale, with consequences that include economic losses and eventual human casualties. Examples abound: among them, denial of service attacks to critical infrastructures have been specially noted worldwide (power grids and nuclear plants), but IoT environments at lower scales are also at risk, including wearables (health information leaks), smart meters (presence at home), connected vehicles, unmanned aerial vehicles and drones and other systems alike [35–37]. Design factors should include a proper risk quantification of the asset to be protected, considering the criticality of the captured data for the operated process or   AQ4 infrastructure, and the degree of resiliency of the asset should such an attack occur. Overall, a thorough understanding of the motivations behind the attacks should be attained to detect recurrent security holes, infer threat patterns and design IoT protocols according to such diagnosed vulnerabilities and thereby prevent future incidences. The fullest potential of IoT systems is strongly subject to its properly protected design against all kind of threats and vulnerabilities, by minimizing the number of security flaws from the very beginning of the design process. If the attack cannot be predicted nor avoided anyhow, to its resilience and capability to recover. To this end, further analyses and efforts must be made to evolve current access control, authentication and identity management mechanisms towards other alternatives encompassing more robust security methods without sacrificing user-friendliness, ease of use and efficiency.

### 1.5.5   Enhanced energy efficiency and autonomy

With the ever-growing scales at which IoT environments are being deployed nowadays in different sectors, there is a remnant question whether current IoT technologies underneath such deployments will meet the aggregate energy demands of IoT meshes. All processing around the IoT node adds up to its computational burden and ultimately, to its autonomy, which is an essential design factor for a number of critical applications (e.g., aerospace or health, among others). As a matter of fact, nowadays no IoT-related study should currently neglect its implications in terms of power consumed at the IoT node, as it has become a decisive factor for its practical adoption. Technologies inherently linked to the batteries themselves such as wireless charging, wake-up circuits, new materials and energy harvesting hardware are nowadays on the focus [38], but the importance of energy efficiency spreads further to the design of the wireless communication stacks, stimulating new wireless interfaces for data broadcasting (e.g., LiFi, Visible Light Communications, VLC), energy-efficient MAC techniques, routing protocols and data collection/compression schemes, among others [39]. This trend hints at a holistic design of the entire IoT application with energy as a strictly necessary design objective at all levels of the communication stack.

Paradoxically in a sense, an emerging market has sprung from the application of IoT networks to the improvement of the energy efficiency of other systems and processes, such as street lighting, home appliances, home automation and heating, ventilation and air conditioning (HVAC) control in buildings, among others [40]. The portfolio of applications under the same motto is increasing day by day, but they occur to be mostly composed of ad hoc solutions with limited interoperability and questionable scalability. A major research effort must be invested along this line, to avoid transforming the global IoT realm into myriads of isolated islands of disconnected functionality and limited exploitable knowledge. By ensuring cross-operations between such islands a competitive advantage can be gained by the simply richer data substrate from which to construct learning models for e.g., temperature prediction or energy consumption forecasting, but also spans business cases around the commercialization of the captured data.

### 1.5.6   (Big) Stream analytics

While this research area can be certainly covered by the aforementioned need for embeddable AI, stream data mining deserves its own space in the IoT technology roadmap. In a word stream mining models (also known as online learning) are subject to stringent storage and processing constraints, calling for customized, computationally lightweight, incremental learning algorithms that do not require the collected data to be stored anyhow [41]. The stream mining scenario may become even more involved when data samples themselves are modelled by nonstationary distributions, a fact that jeopardizes dramatically the discovery of patterns from data. When this is the case, models must not only incrementally build a mathematical model relating their inputs to the target outputs (e.g., IoT sensors inferring how wireless radio measurements relate to the occupancy of the hall), but they must also detect reliably when the discovered relationship between variables change to an extent so as to reactively rebuild the model and ultimately, capture the newly occurring pattern or concept. This paradigm, often referred to as concept drift or evolving concepts, is lately occurring more and more frequently in IoT environments, particularly when sensors and actuators are deployed in scenarios of inherently high dynamics (e.g., industrial plants, smart cities and large events), prone to undergo severe changes in the stationarity of the captured streams. A yet one more degree of complexity is posed by the eventual heterogeneity of such information flows, which has encouraged the adoption of information fusion schemes at different levels of IoT systems [42]. Although notable advances have been made in this area, there are still unaddressed questions in regards to the scalability of practical IoT deployments of stream mining models, since the aggregate demand of the whole IoT mesh is expected to surpass the processing capacity and memory of commodity servers even in cloud infrastructures without latency constraints [43]. To face this challenge, research must be steered towards the distribution of learning tasks all over the mesh, by resorting to distributed computation models as those allowed by software tools (e.g., Spark, Flink, Storm and Samza) or by leveraging concepts from the aforementioned Federated Learning approach.

### 1.5.7    Conclusions

While the above list does not exclude other topics of relevance for IoT systems, these open challenges should lie at the inner core of future research efforts conducted in this area. In essence, IoT systems acquire information about the sensed phenomena and deliver it to smart application and services, which prescribe (to users and/or actuators) how to react according to the captured data. Since IoT communications, protocols and platforms have evolved at a fast pace during the last years, it is not a matter of how the overarching goal of IoT systems will be accomplished, but a matter of when the society will enjoy the benefits of IoT environments as an elemental part of any daily routine.

*Chapter 1*

# The Internet of things: a survey and outlook

## Author Queries

AQ1: Please check the sentence "Because of their nature, though ... value in several IoT scenarios." for clarity.

AQ2: Please check the sentence "On paper, all power saving functionalities ..." for intended meaning.

AQ3: Please check "high-availability" in the sentence "Considering that the upfront costs ..."

AQ4: Is the fragment "**operated process or infrastructure**" fine in the sentence "Design factors should include a proper risk quantification of the asset to be protected, considering the criticality of the captured data for the operated process or infrastructure, and the degree of resiliency of the asset should such an attack occur."?